\documentclass[journal]{IEEEtran}
\IEEEoverridecommandlockouts
\usepackage{cite,subfigure}
\usepackage{amsmath,amssymb,amsfonts,amsthm}
\usepackage{algorithmic}
\usepackage{textcomp}
\usepackage{booktabs} 
\usepackage{makecell} 
\usepackage{xcolor}
\usepackage{pgf,pgfkeys,tikz,circuitikz,float,bm,caption}
\usepackage[ruled,linesnumbered,vlined]{algorithm2e}
\usepackage{graphicx,multirow}%
\usepackage{steinmetz}
\usepackage{siunitx,upgreek}
\def\BibTeX{{\rm B\kern-.05em{\sc i\kern-.025em b}\kern-.08em
		T\kern-.1667em\lower.7ex\hbox{E}\kern-.125emX}}
	
\graphicspath{{figures/}}

\begin{document}

\title{Obfuscation of Human Micro-Doppler Signatures in Passive Wireless RADAR}
\author{Antonios Argyriou \\ 
	Department of Electrical and Computer Engineering, University of Thessaly
	}

\maketitle

\begin{abstract}
		When wireless communication signals impinge on a moving human they are affected by micro-Doppler. A passive receiver of the resulting signals can calculate the spectrogram that produces different signatures depending on the human activity. This constitutes a significant privacy breach when the human is unaware of it. This paper presents a methodology for preventing this when we want to do so by injecting into the transmitted signal frequency variations that obfuscate the micro-Doppler signature. We assume a system that uses orthogonal frequency division multiplexing (OFDM) and a passive receiver that estimates the spectrogram based on the instantaneous channel state information (CSI). We analyze the impact of our approach on the received signal and we propose two strategies that do not affect the demodulation of the digital communication signal at the intended receiver. To evaluate the performance of our approach we use an IEEE 802.11-based OFDM system and realistic human signal reflection models.
\end{abstract}


\section{Introduction}
\label{section:introduction}
Systems that use wireless WiFi signals for passive tracking, localization, and activity detection have been on the rise for several years~\cite{WiVi13,Dai13,wisee13,Ali15,Li16,Wang14,WiHear14,Chen15,jnl_2019_hindawi,cnf_2022_array}. The core idea of these systems is the extraction of the micro-Doppler effects from signals that have impinged on a human. There are two classes of passive techniques that can extract micro-Doppler in WiFi systems, namely Channel
State Information (CSI) based systems, and Passive WiFi Radar (PWR) based systems. What we argue in this paper is that these systems, although extremely useful in a plethora of applications, may allow attacks on human privacy from random wireless receivers of opportunity that can detect human activities when we do not want them to do so. To this aim we study CSI-based systems (since they are more dangerous as we will see) and propose an approach that disables the correct extraction of micro-Doppler effects at a receiver when the transmitter desires to do so.

\begin{figure}[t]
	\centering
	\includegraphics[width=0.95\linewidth]{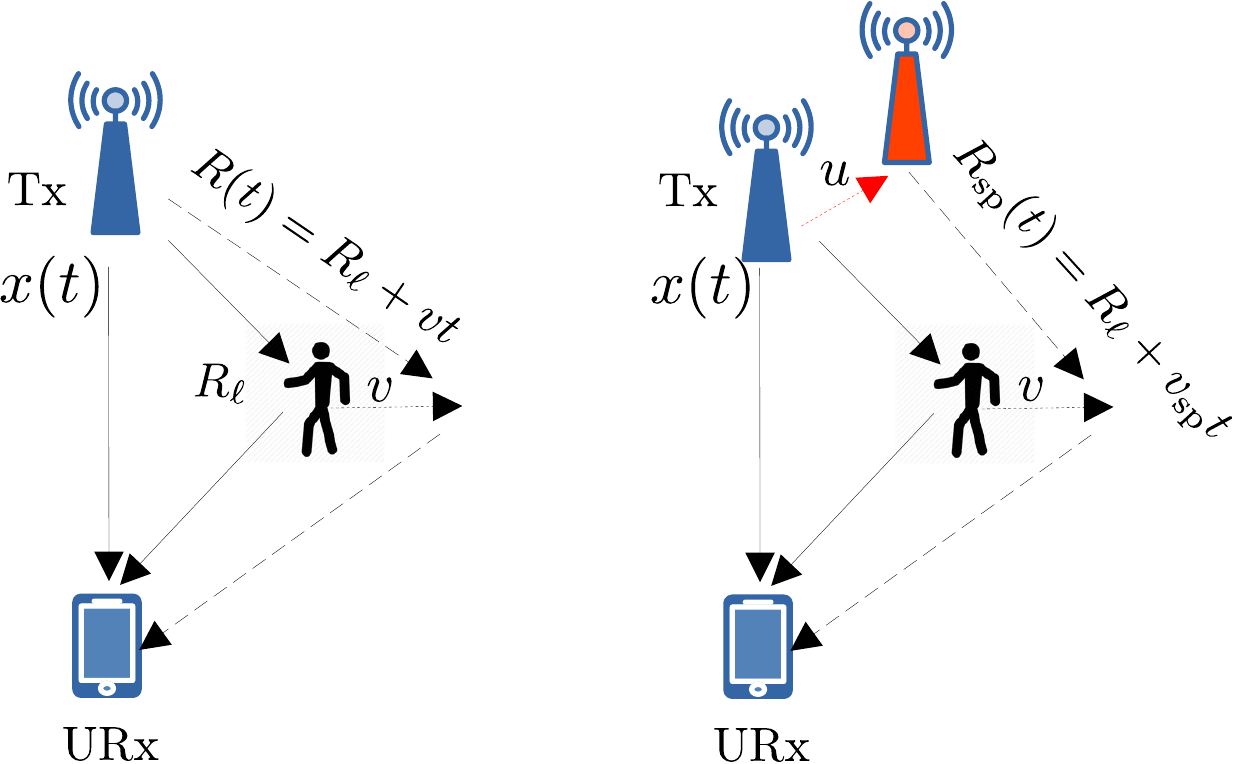}
	\caption{Left: The human moves with a speed that generates a change in the path length equal to $vt$. Right: With the proposed method the spoofed path length (as it appears in the signal) is equal to $(v+v_\text{sp})t$ and is mimicking the effect of a physical movement of the transmitter to the red location.}
	\label{fig:topology-reflections}
\end{figure}

Most PWR systems require a transmitter under the control of the system, i.e. one that it is cooperative and shares a dedicated channel for providing a reference signal to the passive receiver~\cite{Li20} (Fig.~\ref{fig:topology-reflections} left). Other PWR systems like the works in~\cite{WiTrack14,Xie19} leverage a reflected signal from a stationary target as a reference waveform but the human activity classification performance is suboptimal when compared to the case that a dedicated reference channel is present. In the previous PWR systems micro-Doppler is estimated by correlating the received signal with the reference signal, and then identifying when this correlation peaks for different candidate Doppler values. More recent works on PWR systems do not require a reflected signal as a reference, but simply strip away from the transmitted signal any data-dependent phase/frequency shifts leaving only Doppler-induced frequency variations~\cite{cnf_2022_sam}. This leads to improved Doppler estimates~\cite{cnf_2022_sam}. Overall we notice that PWR systems are more robust when there is cooperation with the transmitter  of the signal. This is something that can be clearly avoided if we want to protect the privacy of users moving around a WiFi access point (AP), making thus PWR systems more safe.

Contrary to PWR systems, CSI-based systems~\cite{Li20,Wang15} are potentially more dangerous. In this case the receiver uses CSI estimates for calculating the signal spectrogram allowing thus the identification of unique micro-Doppler signatures. This type of systems might face more difficulties in accurate micro-Doppler estimation, but pose a serious privacy threat since an arbitrary unauthorized user can deploy them without requiring explicit authorization from the AP that is associated with. The basic idea, described in works such as~\cite{Wang15,Zhang19}, is based on the fact that human movement generates variations in the length of the multiple reflected paths from the signal source to the receiver, and consequently the power of the received signal over certain frequency bands (i.e. the channel frequency response (CFR) $H(f,t)$).
As seen in Fig.~\ref{fig:topology-reflections} (left) the length of the $\ell$-th path changes from $R_\ell$ to $R_\ell+vt$) upon human movement.

It is important to note that this class of techniques is not viewed as threatening to the user but as a low cost approach that enables several applications. Hence, there is little focus on combating these passive systems. The few existing systems in the literature focus on obfuscating signal parameters not related to micro-Doppler. As an example friendly Cryptojam~\cite{Rahbari14} prevents estimation of various PHY layer parameter like the modulation type but not that of Doppler, and of course not the micro-Doppler in reflected signals from humans. Artificial frequency shifts in a transmitted signal have been proposed in the past as a method to prevent correct demodulation by an adversary~\cite{jnl_2020_phy}, but again this method was not intended to prevent estimation of micro-Doppler signatures. The most closely related work to this paper may be PhyCloak~\cite{PhyCloak16} that distorts Doppler information at a passive receiver through a third relay node that emits simultaneously with the transmitter. However, this system requires always the presence of a third relay node that is synchronized to the transmitter, while more importantly is not concerned with reflected signals from humans that are affected by micro-Doppler.

As we already stated, in this paper we take a different stance and we argue that these systems can be a threat to privacy. In cases where we do not desire their operation, we should try to prevent them by being used from malicious wireless users. Consequently, we focus on CSI-based schemes where a receiver of opportunity, namely an unauthorized receiver (URx), can calculate the CSI and its frequency-dependent short time scale variations that allow micro-Doppler estimates through a spectrogram. We consider a WiFi transmitter that uses orthogonal frequency division multiplexing (OFDM) and emits a digital communication signal to a desired receiver but it does not want this signal to be used for micro-Doppler estimation by others. \textit{Under this setup we propose to protect humans from disclosure of their micro-Doppler signature by inserting an artificial frequency variation effect in the transmitted signal so as to smear real micro-Doppler effects. At the same time the artificial signal should not affect signal demodulation.} We accomplish that by injecting either a low frequency modulated (FM) signal in the transmitted waveform that smears the micro-Doppler signature, or a signal that spoofs the \textit{path length changes} that take place in a micro scale as illustrated in Fig.~\ref{fig:topology-reflections} and discussed in \cite{Wang15}. We study the received signal and the CSI estimation process at the receiver, and then we investigate the effect of the two aforementioned signal structures on the spectrogram.

\section{Signal Model}
\label{section:signal-model}
For the purpose of this paper we assume that the transmitter (Tx) is static, i.e. it is a base station that is part of the legitimate network with which it communicates (we do not study it). The signal model is that of 802.11a/g/n/ac OFDM where a \SI{20}{\mega\hertz} channel is used~\cite{80211ac}. A moving transmitter can also be supported but the resulting expressions become more convolved without adding any new insights to the core idea of this paper. The 802.11 PHY frame preamble details can be found here~\cite{80211ac}. The basic aspects from the 802.11 frame structure that we need to know for this paper is that the preamble of each frame consists of 7 short and repeating known QAM symbols, as it is typically done in the preambles of wireless frames. These known symbol values allow for CSI estimation at a receiver with a process that we will discuss later.

The one-way delay of the WiFi OFDM signal that impinges on the human target is time-varying and equal to 
\begin{align}
	\tau_\ell(t)=\frac{R_\ell(t)}{c}=\frac{R_\ell}{c}-\frac{v_\ell t}{c}, 
	\label{eqn:tau_ell}
\end{align}
where $R_\ell$ and $R_\ell(t)$ correspond to the length of the $\ell$-th path before the movement and at time $t$ respectively (see Fig.~\ref{fig:topology-reflections}). $v_\ell $ is a speed that results in the so called \textit{path length change}~\cite{Wang15} of $v_\ell t$ in time $t$. Since different body parts travel different distances and have different speed as the human moves, this results in a unique micro-Doppler signature when the effect of all the paths is aggregated.

We will consider the impact of micro-Doppler when the transmitted signal $x(t)$ is the result of multi-carrier modulation and more specifically OFDM. With $N$ subcarriers that are spaced relative to the carrier $f_c$ at locations $f_k$=$k \Delta f$ Hz that can contain data, pilot symbols, or a combination of both (depending on the standard), the desired baseband OFDM symbol in continuous time is:
\begin{align}
	x(t)=\frac{1}{\sqrt{N}}\sum_{k=0}^{N-1}X[k]e^{j2\pi k \Delta f t},~~0\leq t \leq T_N
	\label{eqn:ct-ofdm} 
\end{align}
$X[k]$ is the complex QAM symbol modulated onto subcarrier $k$, and $T_N=N/\Delta f$ is the OFDM symbol duration. 

The channel transfer function or channel frequency response (CFR) of a wideband time-varying channel over all the paths is
\begin{align}
	H(f,t)=\sum_\ell h_\ell(t) e^{-j2\pi f \tau_\ell(t)},
	\label{eqn:frequency-response}
\end{align}
where $h_\ell(t)$ is the complex channel gain of the $\ell$-th path excluding Doppler.  The channel impulse response at time $t$ is time-varying (p.p. 21~\cite{tse}):
\begin{align}
	h(\tau,t)=\sum_{\ell} h_\ell(t) \delta(\tau-\tau_\ell(t)) 
	\label{eqn:impulse-response}
\end{align}
To derive the received signal we have to combine~\eqref{eqn:ct-ofdm},~\eqref{eqn:frequency-response} and \eqref{eqn:impulse-response}. At the passive URx the received baseband continuous time signal after transmission in a time-varying frequency selective channel is the convolution of the channel impulse response with the transmitted OFDM signal. We also include the additive white Gaussian noise (AWGN) $w(t)$ and obtain:
\begin{align}
	y(t)&=\int_{-\infty}^{\infty}h(\tau,t)x(t-\tau)d\tau+w(t)\nonumber\\
	&=\sum_{\ell} h_\ell(t) x(t-\tau_\ell(t))+w(t)\nonumber\\
	&=\frac{1}{\sqrt{N}}\sum_{k=0}^{N-1} \sum_{\ell}^{} h_\ell(t) X[k] e^{j2\pi k\Delta f(t-\tau_\ell(t))}+w(t)\nonumber\\
	&=\frac{1}{\sqrt{N}}\sum_{k=0}^{N-1}X[k]e^{j2\pi k\Delta ft }\sum_{\ell}^{}h_\ell(t) e^{-j2\pi k\Delta f\tau_\ell(t)}\nonumber\\
	&+w(t) \label{eqn:signal-model-with-ofdm-cfr0}\\
	&=\frac{1}{\sqrt{N}}\sum_{k=0}^{N-1}X[k]e^{j2\pi k\Delta ft }H(k\Delta f,t)+w(t)\label{eqn:signal-model-with-ofdm-cfr}
\end{align}
By replacing \eqref{eqn:frequency-response} to \eqref{eqn:signal-model-with-ofdm-cfr0} we obtain \eqref{eqn:signal-model-with-ofdm-cfr} which is another form of the output signal that we will use. When the CFR is not time-varying by applying DFT to~\eqref{eqn:signal-model-with-ofdm-cfr0},~\eqref{eqn:signal-model-with-ofdm-cfr}  we get the well known result of OFDM where the channel becomes from a frequency selective fading channel to a flat fading channel, i.e. $Y(k \Delta f)$=$X[k]H(k \Delta f)$. Of course in practice time variations of the CFR cause inter-carrier interference (ICI).

To get another useful expression we can also substitute $\tau_\ell(t)$ from~\eqref{eqn:tau_ell} to~\eqref{eqn:signal-model-with-ofdm-cfr0}. Then the term $e^{-j2\pi k\Delta f \frac{R_\ell}{c}}$, that is constant phase for the $\ell$-th path (i.e. it is not indexed by $t$) can be merged with $h_\ell(t)$ to be jointly denoted as $h_\ell^{'}(t)$. We then have:
\begin{align}
	y(t)& =\frac{1}{\sqrt{N}}\sum_{k=0}^{N-1}X[k]e^{j2\pi k\Delta ft }\sum_{\ell}^{}h^{'}_\ell(t) e^{-j2\pi k\Delta f\frac{v_\ell t}{c}}+w(t) \label{eqn:signal-model-with-ofdm-speed}
\end{align}

Note that we do not model a carrier frequency offset (CFO) since we have seen that it affects marginally the resulting spectrograms. OFDM signal models that include CFO and artificial frequency variations in the transmitter have been developed in~\cite{2022-argyriou-arxiv} and can be combined with the ideas in this paper if needed. We also ignore the sampling clock offset (SCO) at the passive URx since we are not interested in its ability to correctly demodulate OFDM symbols (by sampling when the matched filter output peaks).\footnote{To see how this can be solved for OFDM we refer the reader to related works on the topic, e.g. see~\cite{yip04}.} 

\section{Micro-Doppler Smearing \& Spoofing}
We can design an artificial signal to have any desired form as long as it does not compromise decoding of $x(t)$. We denote this signal as $x_\text{sp}(t)$ in the time domain and $X_\text{sp}(f,t)$ is the Fourier Transform (FT). The two ideas that we investigate are: \textit{Method 1, subcarrier independent micro-Doppler obfuscation with spectrogram smearing. Method 2, subcarrier-dependent spoofing for mimicking path length changes.} We must note that under both scenarios the injected signal does not introduce observable frequency shifts in the scale of an OFDM frame, thus it does not prevent CSI estimation from the nominal receiver that desires to demodulate the transmitted frame. 
IEEE 802.11 frequency estimation and compensation algorithms are robust to Doppler variations of a few hundred \SI{}{\kilo\hertz}, which is far beyond what we are doing with the artificial signal in this work~\cite{Wang18}. The OFDM BER performance under this range of artificial frequency variations was studied in~\cite{cnf_2023_radarconf1} and no performance degradation was observed.

\subsection{Method 1: Smearing}
With the first method we introduce an oscillating sinewave with a frequency of a few 10's of \SI{}{\hertz} in the transmitted signal $x(t)$ to smear the micro-Doppler effects of human movement that are in this range~\cite{Vishwakarma21}. Note that we do not have to re-create a specific spectrogram but we only have to destroy the signature that reveals micro-Doppler dependence to different user movement speeds. 

To generate the final signal at the transmitter we re-modulate the information signal which consists of the QAM symbols $X[k]$ before IDFT, with a frequency domain signal $X_\text{sp}(f,t)$. In the time domain this is defined to be an frequency-modulated (FM) waveform with maximum instantaneous frequency shift $\delta f$, and frequency $f_m$, making thus the instantaneous frequency equal $f_i(t)$=$f_c+\delta f \cos (2\pi f_m t)$~\cite{proakis02}. This makes the actual time-domain FM \textit{smearing signal} to be:
\begin{align}
	x_\text{sp}(t)=e^{j \frac{\delta f}{f_m}\sin(2\pi  f_m t)}
	\label{eqn:ftg-signal}	
\end{align}
Hence, this waveform produces a signal that in the frequency domain speads between the maximum instantaneous frequencies $-\delta f$ and $+\delta f$, at a rate of $f_m$ \SI{}{\hertz}. The result is that it smears the frequency domain micro-Doppler signature in that frequency range. An important detail from the literature is that the frequency at which the various human body parts move is very small and in the range of a few \SI{}{\hertz}~\cite{Boulic90,Vishwakarma21}. Consequently, the variation of the frequency $f_m$ should be higher that this value which means that just a few tens of~\SI{}{\hertz} for $f_m$ will smear very well the received signal spectrogram.

\subsection{Method 2: Spoofing Path Length Changes} The second approach is to think of what would the signal look like if the transmitter had actually moved. Hence, this technique can be more appropriately referred to as \textit{spoofing} and not simply obfuscation. Note that we can only spoof the phase change in a way that mimicks the signal that would be produced if a physical movement of the transmitter took place (to the "new" red location in Fig.~\ref{fig:topology-reflections}(right)). The received signal has to be such that the path length change seen in Fig.~\ref{fig:topology-reflections} will not be $v_\ell t$ but $v^{'}_\ell t$ in the same time period of $t$ seconds, that is we want the URx to receive the following signal:
\begin{align}
	y(t)=\sum_{k=0}^{N-1}\frac{X[k]}{\sqrt{N}}e^{j2\pi k\Delta ft }\sum_{\ell}^{}h^{'}_\ell(t) e^{-j2\pi k\Delta f\frac{v^{'}_\ell }{c}t} 
	\label{eqn:signal-model-with-spoofing-path-length-changes}
\end{align}
This is a form of the signal that we would like to be received at the URx and the question is how to create it. To obtain the above we set the \textit{spoofing signal} equal to $x^{(k)}_\text{sp}(t)=e^{j 2 \pi k\Delta f \frac{v_\text{sp}}{c}t} $, making thus \eqref{eqn:signal-model-with-ofdm-speed}:
\begin{align}
	y(t)&=\sum_{k=0}^{N-1}\frac{X[k]x^{(k)}_\text{sp}(t)}{\sqrt{N}}e^{j2\pi k\Delta ft }\sum_{\ell}^{}h^{'}_\ell(t) e^{-j2\pi k\Delta f\frac{v_\ell}{c}t}\nonumber\\
	&=\sum_{k=0}^{N-1}\frac{X[k]e^{j 2 \pi k\Delta f \frac{v_\text{sp}}{c}t}}{\sqrt{N}}e^{j2\pi k\Delta ft }\sum_{\ell}^{}h^{'}_\ell(t) e^{-j2\pi k\Delta f\frac{v_\ell}{c}t}\nonumber\\
	&=\sum_{k=0}^{N-1}\frac{X[k]}{\sqrt{N}}e^{j2\pi k\Delta ft }\sum_{\ell}^{}h^{'}_\ell(t) e^{-j2\pi k\Delta f\frac{v_\ell-v_\text{sp}}{c}t} 
	\label{eqn:signal-model-with-spoofing-path-length-changes2}
\end{align}
By comparing \eqref{eqn:signal-model-with-spoofing-path-length-changes} and \eqref{eqn:signal-model-with-spoofing-path-length-changes2} we see that the path length change that will be perceived for the $\ell$-th path is $v^{'}_\ell =v_\ell-v_\text{sp}$. So in the practical implementation instead of sending to each subcarrier complex symbol $X[k]$ we send $X[k]x^{(k)}_\text{sp}(t)$, that is a subcarrier and time dependent term. In the next section we discuss how these two methods will affect the CFR power estimate that will be used for spectrogram estimation by the URx.

\section{Impact of CFR Power}
The CFR power for subcarrier $k$ is estimated after we first match filter the frequency domain received signal of that subcarrier, namely $Y(k\Delta f)$, with the known preamble $X[k]$.  Since the spoofing signal has a FT $X_\text{sp}(f,t)$ this leads to the CFR being equal to:
\begin{align}
	\Big | \frac{X^*[k] Y(k\Delta f)}{|X[k]|^2} \Big |^2=| X_\text{sp}(k\Delta f,t) H(k \Delta f,t) |^2 \label{eqn:cfr-power-estimate}
\end{align}
From the above it is clear that we can select a smearing or spoofing waveform so that it alters completely the CFR power estimate above as a function of frequency or time.

\textbf{Method 1:} With this method the result is straightforward since the signal is independent of subcarriers. This method simply smears the resulting signal over the specified bandwidth of the FM signal obfuscating thus micro-Doppler effects. This bandwidth is approximated to be $2(\frac{\delta f}{f_m}+1)$ according to Carson's rule.

\textbf{Method 2:} Now we analyze the CFR power, namely $| X_\text{sp}(k\Delta f,t)H(f,t)|^2$, to understand what takes place. An expression for the CFR power $|H(f,t)|^2$ has been studied in~\cite{Wang15}, but as we have seen in~\eqref{eqn:signal-model-with-spoofing-path-length-changes} the path length is now affected by the artificial speed $v_\text{sp}$. Hence, we re-calculate the CFR power as in~\cite{Wang15} that now becomes for the set of all end-to-end paths $\mathcal{P}$:
\begin{align}
	& | X_\text{sp}(k\Delta f,t) H(k \Delta f,t) |^2 = \sum_{\ell \in \mathcal{P}} D_1(f,t) \cos (\frac{2\pi v^{'}_\ell t}{c}+\frac{2\pi R_\ell }{c})  \nonumber \\
	&+ \sum_{\ell,m \in \mathcal{P},\ell \neq m} D_2(f,t) \cos (\frac{2\pi (v_\ell-v_m)  t}{c}+\frac{2\pi (R_\ell-R_m }{c})  \nonumber \\
	&+\sum_{\ell \in \mathcal{P}} D_3(f,t) 
\end{align}
The terms $D_1, D_2, D_3$ are time and frequency dependent and are independent of the path length changes expressed through the speed $v_\ell$. From this expression we see that the CFR power is a function of the actual path length change $v_\ell-v_m$, that are not affected by this type of spoofing, but there are also sinusoidal terms that are affected (in the first summation). The net result is that the correct overall CFR power function is still invalidated by this approach. As we will soon see it has a completely different impact on the Doppler spectrogram than method 1.

\begin{figure}[t]
	\centering
	\subfigure[Pedestrian speed 0.8m/s]{\includegraphics[width=0.499\linewidth]{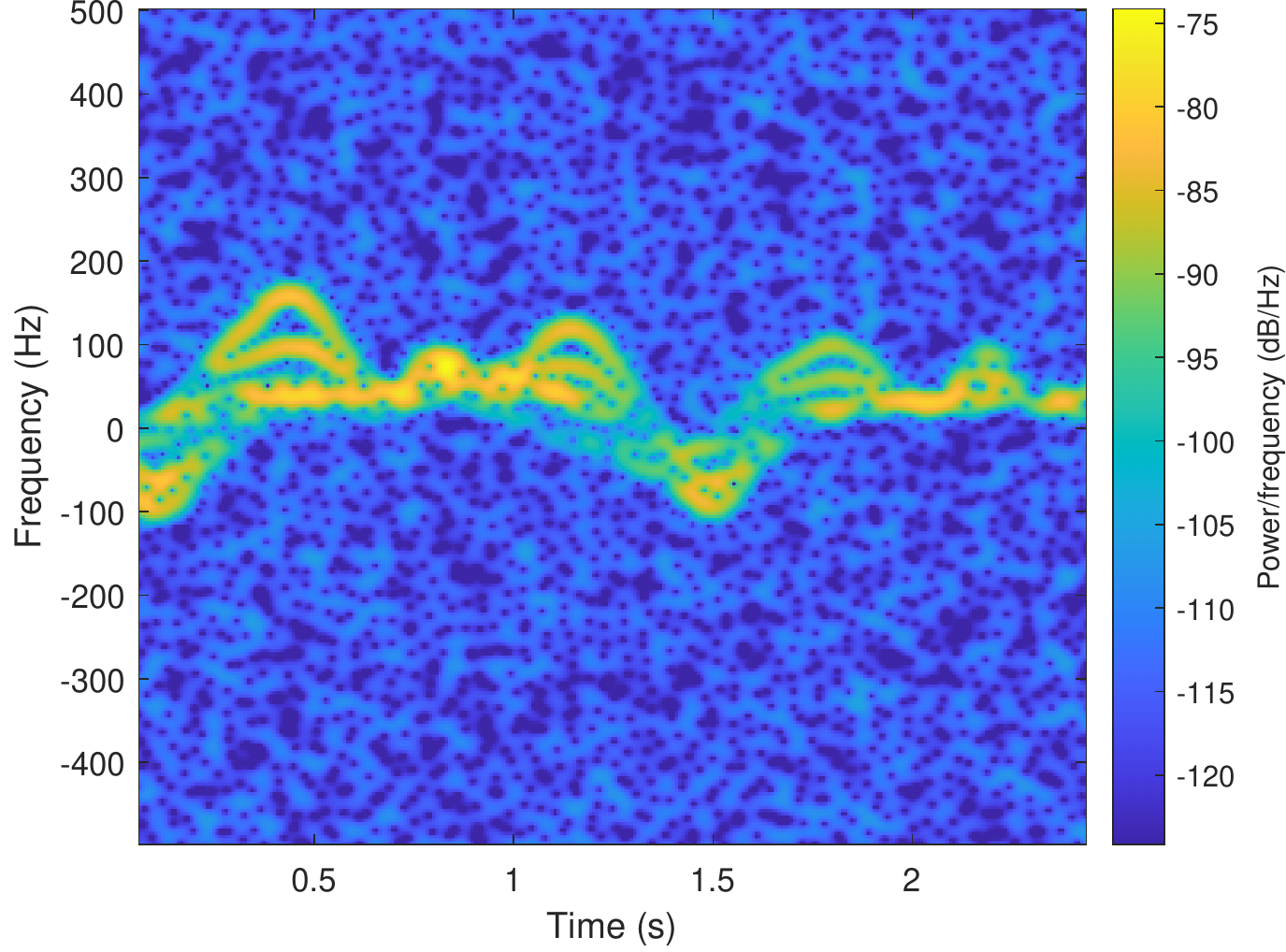}
		\label{fig:speed0.8}
	}
	\subfigure[Pedestrian speed 1.5m/s]{\includegraphics[width=0.499\linewidth]{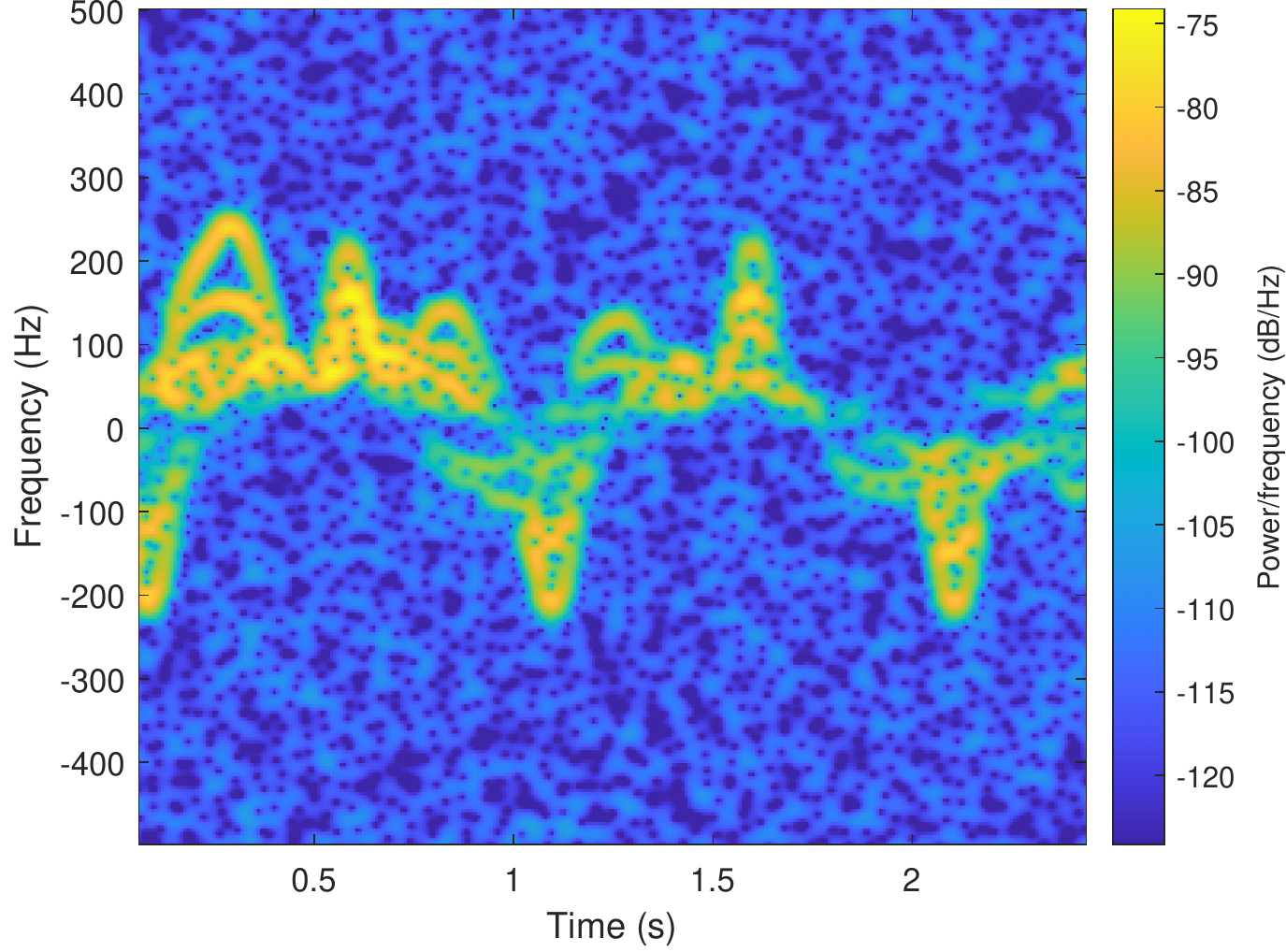}
		\label{fig:speed1.5}
	}
	\caption{Spectrograms of the received signal (before OFDM DFT) for different pedestrian speeds without any form of smearing.}
	\label{fig:radar-ofdm-range-doppler0}
\end{figure}

\begin{figure}[t]		
	\subfigure[Pedestrian speed 0.8m/s w/ smearing under method 1]{\includegraphics[width=0.499\linewidth]{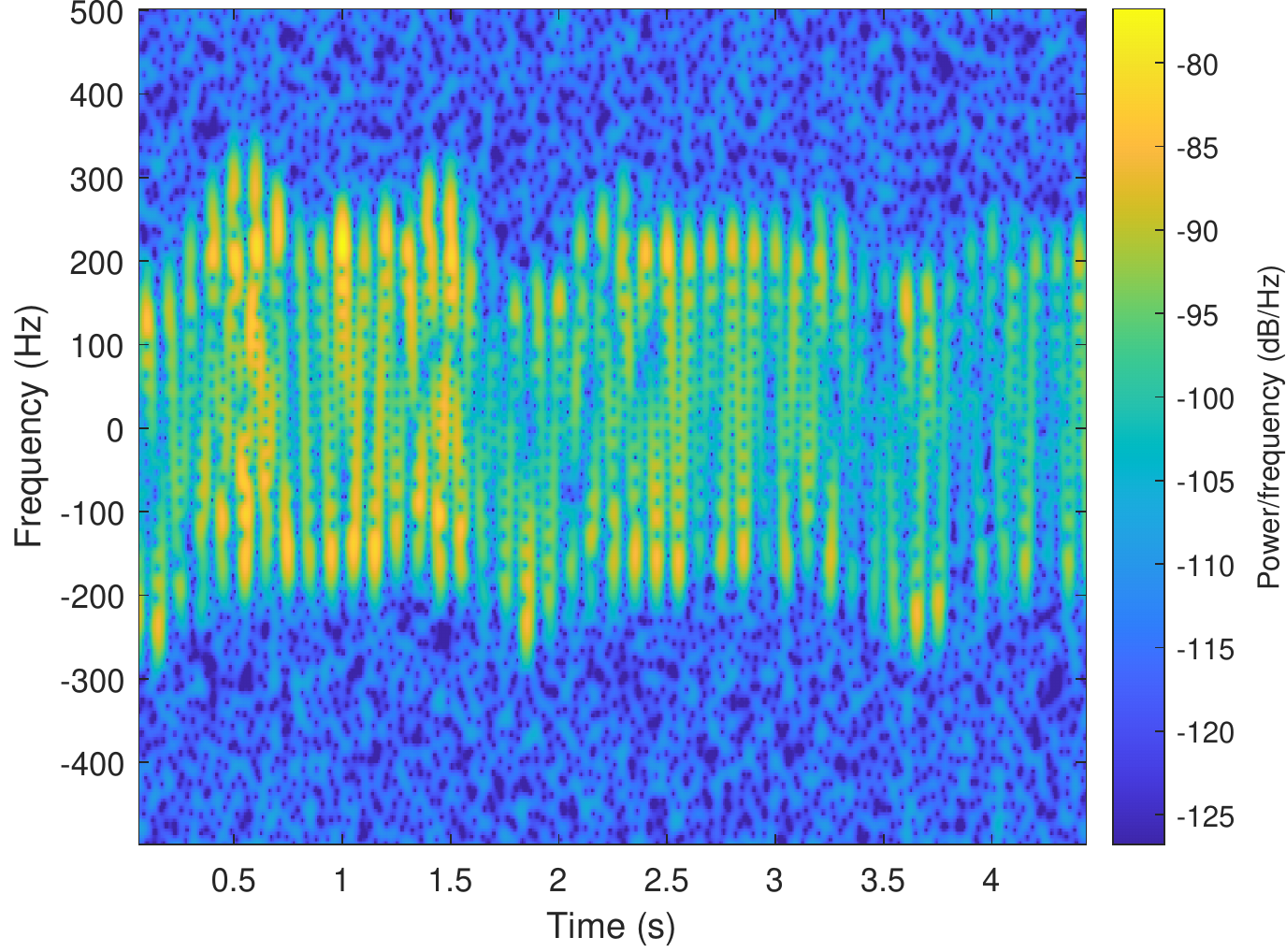}
		\label{fig:speed0.5_Df100_fm10}
	}
	\subfigure[Pedestrian speed 1.5m/s w/ smearing under method 1]{\includegraphics[width=0.499\linewidth]{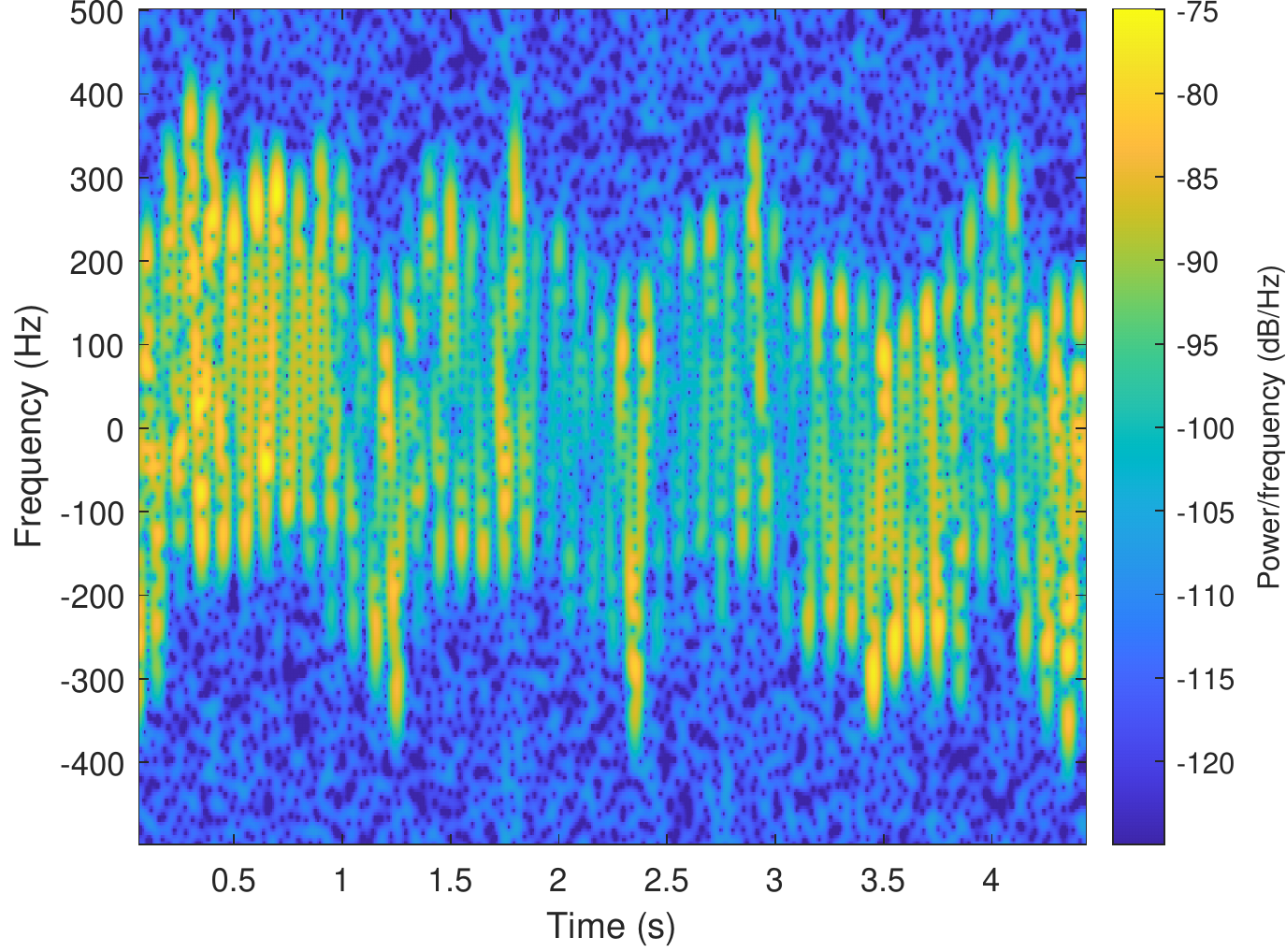}
		\label{fig:speed1.2_Df100_fm10}
	}
	\caption{Spectrograms of the received signal (before OFDM DFT) for different pedestrian speeds and smearing under method 1.}
	\label{fig:radar-ofdm-range-doppler}
\end{figure} 

\begin{figure*}[!htb]
	\centering
	\subfigure[Pedestrian speed 0.8m/s]{\includegraphics[width=0.327\linewidth]{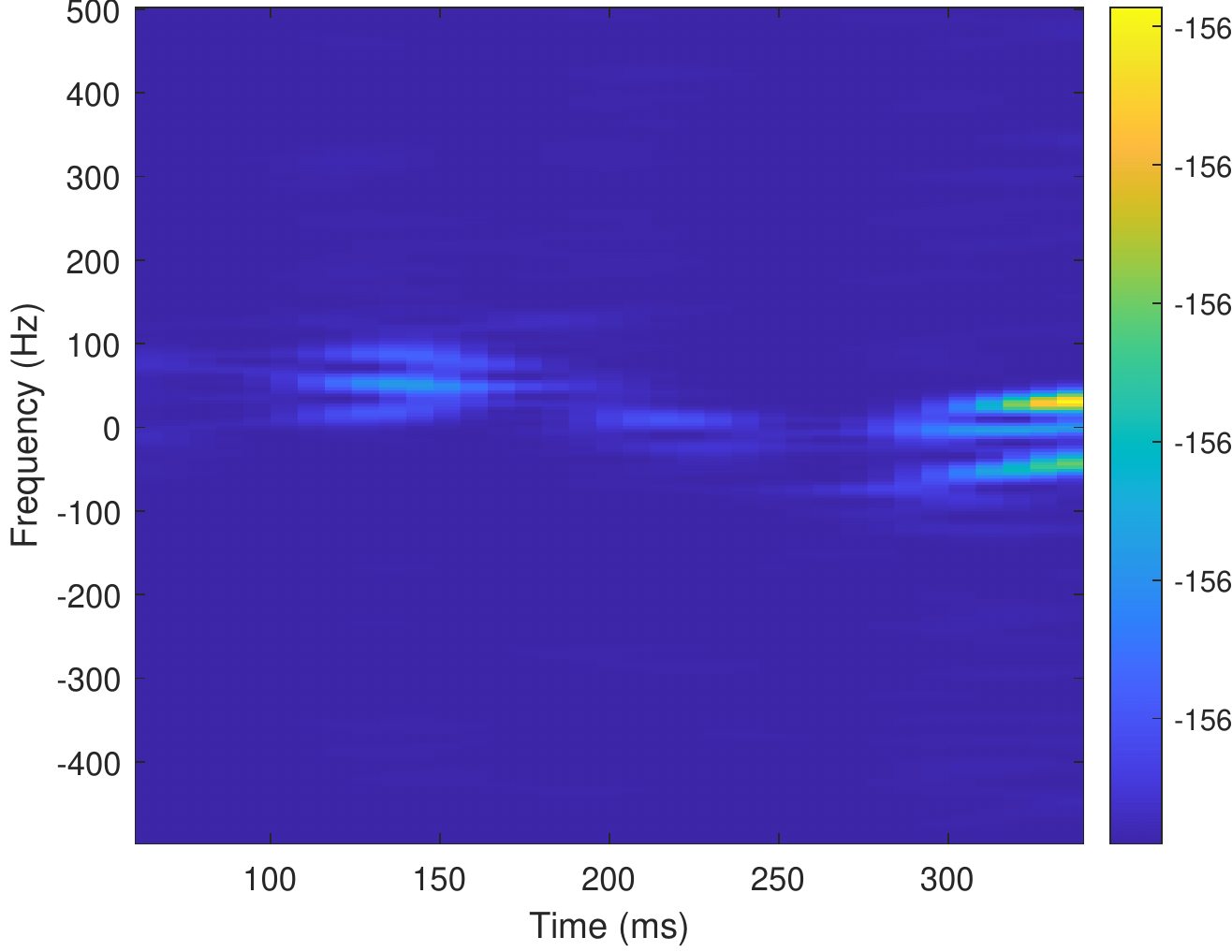}
		\label{fig:ofdm_nosp_cfr_speed.0.8}
	}\hspace{-0.18cm}
	\subfigure[Pedestrian speed 1.5m/s]{\includegraphics[width=0.327\linewidth]{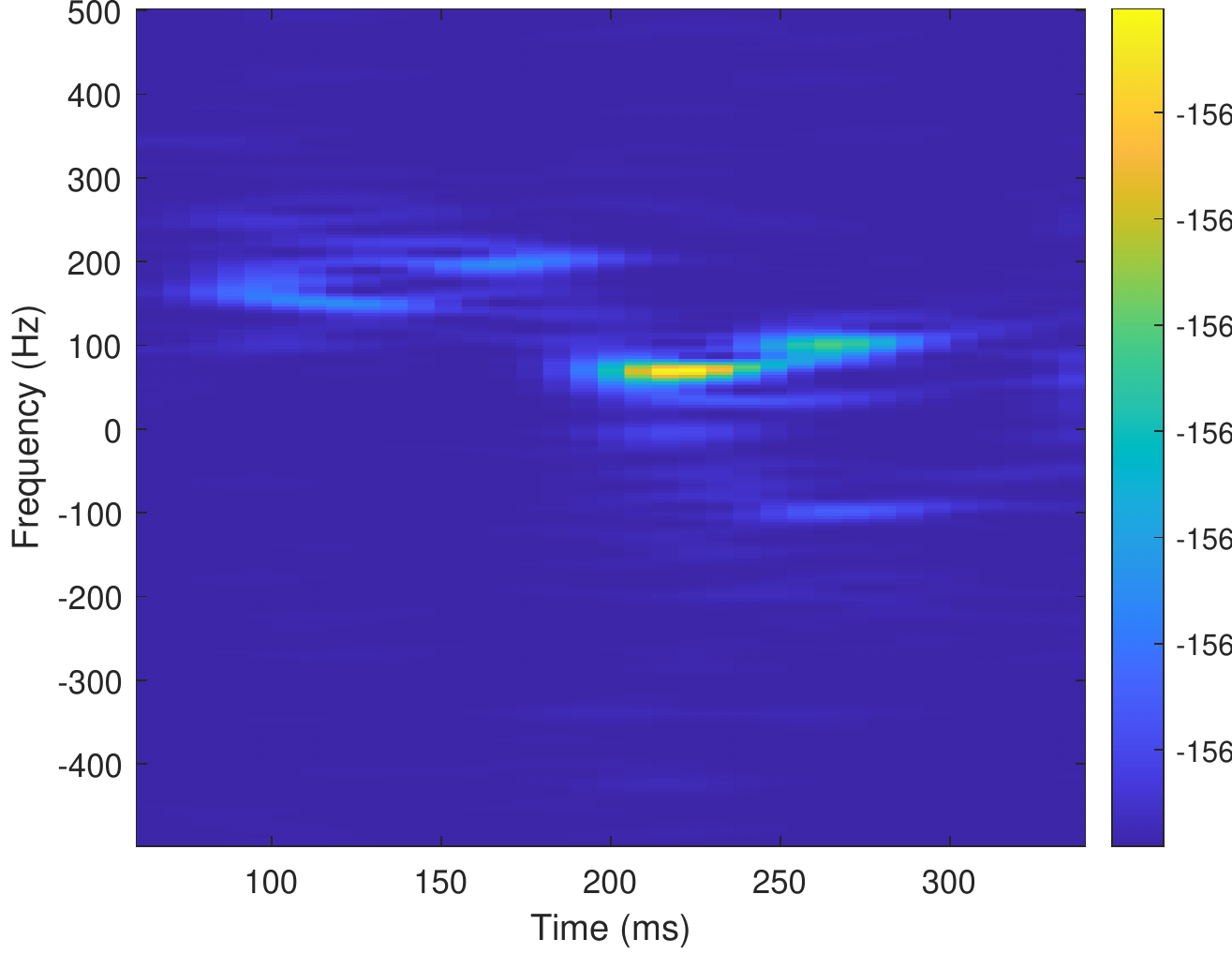}
		\label{fig:ofdm_nosp_cfr_speed.1.8}
	}
	\hspace{-0.18cm}
	\subfigure[Pedestrian speed 0.8m/s \& spoofing]{\includegraphics[width=0.327\linewidth]{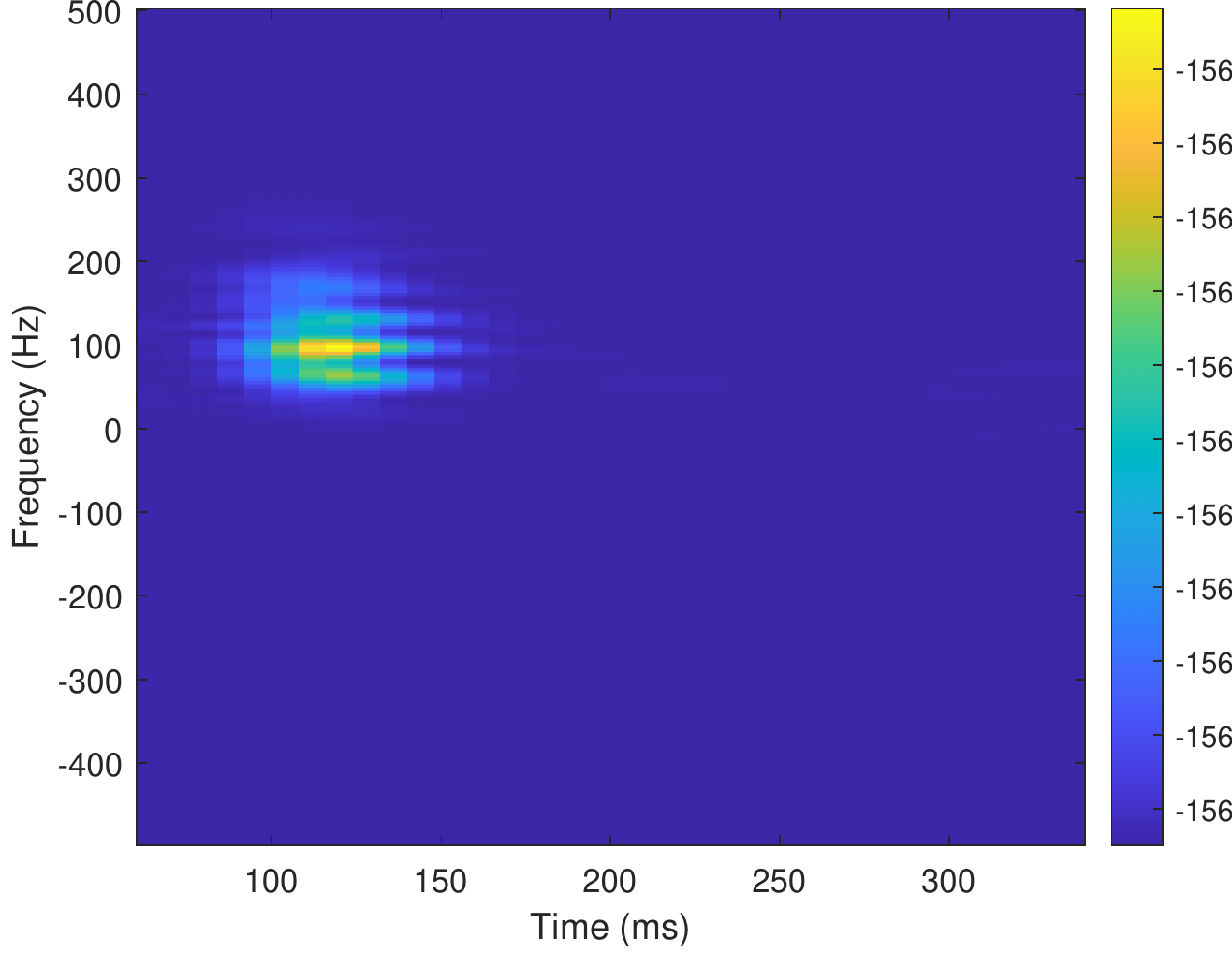}
		\label{fig:ofdm_sp2_cfr_speed.0.8_vsp16}
	}
	\caption{Spectrograms of the CFR power in~\eqref{eqn:cfr-power-estimate} for different pedestrian speeds without spoofing and with micro-Doppler spoofing under method 2.}
\end{figure*} 

\section{Results}
The objective of our simulations is to evaluate the ability of the two methods to smear the Doppler signature of the human movement. We considered an 802.11-based OFDM system that uses a \SI{20}{\mega\hertz} channel, and 64 subcarriers spaced at $\Delta f$=\SI{312.5}{\kilo\hertz} apart. 52 are data subcarriers while 4 subcarriers are used for pilot signals. 
We generate signatures for pedestrians that move at different speeds based on publicly available models in~\cite{Boulic90,Vishwakarma21}. These channel models generate movement of different human body parts and alter the cross-section of the human depending on the movement pattern. 

\subsection{Method 1}
To illustrate better the effects of this method we noticed that is better to calculate the spectrogram on the received signal itself, i.e. before passing the OFDM signal to the DFT for demodulation.
In Fig.~\ref{fig:speed0.8},~\ref{fig:speed1.5} we present spectrogram results for humans moving at different speeds without any form of micro-Doppler smearing. We can easily see the micro-Doppler for the different speeds and the peaks that occur at different frequencies. In Fig.~\ref{fig:speed0.5_Df100_fm10} and Fig.~\ref{fig:speed1.2_Df100_fm10} we present results for a pedestrian moving again at \SI{0.8}{\meter/\second} and \SI{1.5}{\meter/\second} under the smearing method 1. It is clear that the Doppler smearing method spreads the resulting signal over an area of $\delta f$$\approx$\SI{\pm 200}{\hertz} in both cases making thus the spectrograms lack any distinguishing features. $f_m$ which was set to \SI{10}{\hertz} is already high enough to cover the moving user that presents peaks in the spectrograms of Fig.~\ref{fig:speed0.8},~\ref{fig:speed1.5} with at most \SI{0.5}{\second} (\SI{2}{\hertz}). Higher speed user movements will result in more rapid changes over time that can still be covered with an $f_m$ of a few 10's of \SI{}{\hertz}.

\subsection{Method 2}
For the results under method 2 we demodulate each OFDM symbol with DFT, and then we  produce a fine estimate of the time-varying CFR power for each one of the used subcarriers. CFR power is estimated for each one of the subcarriers based on the OFDM symbols in the 802.11 preamble allowing the calculation of $\Big | \frac{X^*[k] Y(k\Delta f)}{|X[k]|^2} \Big |^2$ since the $X[k]$ are known. From this time-varying CFR power estimate we calculate the spectrogram with a resolution of a single OFDM symbol.

The related results for the spectrogram of the CFR power without spoofing can be seen in Fig.~\ref{fig:ofdm_nosp_cfr_speed.0.8}, Fig.~\ref{fig:ofdm_nosp_cfr_speed.1.8}, and with spoofing in Fig.~\ref{fig:ofdm_sp2_cfr_speed.0.8_vsp16}. We can now see that for the same human speed we can generate a different spectrogram at the URx which is more similar to a spectrogram of a human moving at a higher speed. Now there is not a consistent smearing all over the frequency range but a shift in frequency and time of the human signature that depends on the path length change $v^{'}_\ell t$. Different values for $v_\text{sp}$ will result in different placement of the signature in the spectrogram.

\section{Conclusions}
In this paper we presented a new approach protecting humans from disclosing the micro-Doppler signature of their movements in passive wireless RADAR systems. We proposed the insertion of artificial frequency variations in the transmitted signal for smearing the passively estimated spectrogram of the received signal. We investigated two approaches that serve equally well the desired goal. The proposed idea can be used with minimal cost and overhead for protecting the privacy of wireless users from malicious unauthorized wireless users without any impact on digital communication performance.

\bibliographystyle{IEEEtran}
\bibliography{../../../../../tony-bib}

\end{document}